\documentclass[graphicx]{revtex4-1}
\usepackage{graphicx}
\usepackage[cp1251]{inputenc}  
\usepackage[T2A]{fontenc}

\begin{document}

\title{Non-linear additives to the Brooks-Herring screened potential}

\author{P. N. Romanets}

\begin{abstract}
In the present work we calculate the non-linear additives to the Brooks-Herring electrostatic screened potential. We also calculate the corresponding additives to the ionized impurity scattering mobility in n-GaAs over a wide range of doping levels and temperatures. It is demonstrated that the additives to the mobility changes its value up to 50\%. The most dramatically mobility is changed near the Mott transition, when the doping level is about $10^{14}$--$2\times10^{16}$ cm$^{-3}$ and temperature is below 50 K. Also, the results allow us to conclude that the non-linear additives break the cross section symmetry with respect to the sign of the charge of the scattering center.
\end{abstract}

\maketitle

\section{Introduction}

The scattering of electrons by ionized impurities in solids has been studied for more than half a century. The problem is related to the nature of long-range action of the electrostatic potential, which complicates the reduction of the many-body problem to the problem of two bodies. The basic idea for solving the problem is to perform statistical averaging of the field source over the all charged particles. Thus averaging leaves only effective near-field, while the far-field averages to zero. It was suggested many different theoretical models during the last 70 years. The most famous of them are the Conwel and Weisskopf model \cite{Conwell&Weisskopf}, models that reduce the problem to the third body exclusion method\cite{Ridley1,Ridley2,Poznyakov}, Brooks-Herring model \cite{Brooks1,Brooks2}  and model of the partial wave phase shifts \cite{Blatt,Fridel}.

The Conwel and Weisskopf model was a simple attempt to avoid divergence in collusion integral by using cutoff parameter therefore it is not quite consistent.

Model of the partial wave phase shifts has been developed for the case of extremely high carrier concentrations realized in metals. In this case the Born approximation for transition probability is not justified and angular momentum became essential quantum number for the scattering process.

The idea of third body exclusion methods is that scattered by the given center particle must not be scattered by another one. On our opinion, the cross section must not be subjected by this additional condition because it follows from the Boltzmann kinetic equation.

Given the above, the Brooks-Herring model is the most consistent for low and moderate carriers concentrations (criterion can be found in \S 9, ref. \cite{Gantmakher&Levinson}).

 The validity of the Brooks-Herring approach was discussed  several times \cite{Gantmakher&Levinson,Falicov&Cuevas,Chattopadhyay&Queisser}, but to the best of our knowledge  this investigation mathematically not rigorous or just comparable. Moreover, previous authors consider only limiting cases of non-degenerate or degenerate electron gases.

 It is well known that Brooks-Herring screened potential $\phi_1(r)\propto\exp(-r/\lambda)/r$  can be obtained as the solution of the Poisson equation with linearized right-hand side. The parameter $\lambda$ is named screening length.

 The main issues that arise in this model is the validity of the expansion of the right-hand side of Poisson's equation in power series and accuracy which implements a linear term of the expansion.

In the present investigations, we obtain the approximate solution of Poisson's equation that contains the following non-linear terms of the expansion. As an example, the calculation of the ionized impurity scattering mobility is performed for n-GaAs parameters. The results of the calculation give an opportunity to determine which areas and with what precision Brooks-Herring approach is valid. In addition, the results also provide an opportunity to assess the accuracy of the formulas for mobility with the following non-linear terms.

\section{Screened potential}
 The Poisson equation for the screened potential has the form
 \begin{eqnarray}
 \Delta \phi({\bf r})=\frac{4\pi e}{\epsilon}\left\{n[\zeta-e\phi({\bf r}),T]-n[\zeta,T]\right\},\label{eq1}
 \end{eqnarray}
 where $n(\zeta,T)=2[mT/(2\pi\hbar^2)]^{3/2}{\cal F}_{1/2}(\zeta/T)$ is the electron concentration as the function of temperature $T$ and chemical potential $\zeta$, ${\cal F}_{1/2}(x)$ is the Fermi-Dirac integral \cite{Seeger}, $m$ is the effective mass, $\epsilon$ is the static dielectric constant and $e=-|e|$ is the electron charge.
 We will restrict the investigation to the centrally symmetric case for the bulk semiconductor electron gas ($\phi({\bf r})\equiv \phi(r)$). Leaving only the first non-zero term of the right-hand side power-series expansion, one can obtain the Brooks-Herring approach for the screened potential $\phi_1(r)$ with the screening length $\lambda$:
 \begin{eqnarray}
 \lambda^2=\frac{4\pi e^2}{\epsilon}\frac{\partial n(\zeta,T)}{\partial \zeta}.\label{eq2}
 \end{eqnarray}
 Below we use dimensionless coordinate $x=r/\lambda$ and potential $\phi(r)=\varphi(x)Z|e|/\epsilon\lambda$, where $Z$ is the charge  of the scattering center in $|e|$ units.  Preceding the expansion of the right-hand side of eq. (\ref{eq1}) one obtains:
 \begin{eqnarray}
 \frac{d^2 \varphi(x)}{dx^2}+\frac{2}{x}\frac{d\varphi(x)}{dx}-\varphi(x)=\sum_{l=2}^{L}\gamma_{l} \varphi(x)^l+R_L(x), \qquad \gamma_{l}=\frac1{l!}\frac{\frac{\partial^l n(\zeta,T)}{\partial \zeta^l}}{\frac{\partial n(\zeta,T)}{\partial \zeta}}\left[\frac{Ze^2}{\epsilon\lambda}\right]^{l-1},\label{eq3}\nonumber\\
 R_L(x)=\frac1{(L+1)!}\frac{\frac{\partial^{L+1} n[\zeta+e\xi\phi(x),T]}{\partial \zeta^L}}{\frac{\partial n(\zeta,T)}{\partial \zeta}}\left[\frac{Ze^2}{\epsilon\lambda}\right]^{L},\qquad 0<\xi<1;
 \end{eqnarray}
 where we use the remainder term in the Lagrange form \cite{Taylor_Klein}. Note that since the functions ${\partial^{L+1} n(\zeta,T)}/{\partial \zeta^{L+1}}$ have a maximum in the region $\zeta\in(-T,~T)$, in the degenerate case strict inequality holds $|R_L(x)|<|\gamma_{L+1}\phi(x)^{L+1}|$ ($Z>0$), whereas in the case of Boltzmann electron gas we can speak only about the order of magnitude $|R_L(x)|\sim|\gamma_{L+1}\varphi(x)^{L+1}|$. We proceed the consideration, omitting the remainder term $R_L(x)$ and supposing that tolerance is estimated by the value $|\gamma_{L+1}\varphi(x)^{L+1}|$.

\begin{figure}
\centering
\includegraphics[width=3in]{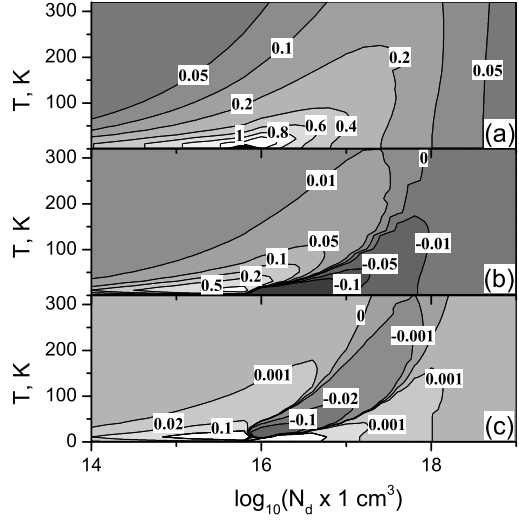}
\caption{The coefficients $\{\gamma_{l}\}$ for n-GaAs parameters versus temperature $T$ and donor concentration $N_d$ (logarithmic scale) as the contour plots. Frames a-c correspond to $\gamma_{2-4}$, $Z=1$. \label{Fig1}}
\end{figure}

Supposing that $\{\gamma_l\}$ are low parameters one can  replace $\varphi(x)$ with $\varphi_0(x)=\beta_0\exp(-\alpha_0x)/x$ in the right hand side. The latter designation contains  dimensionless parameters $\beta_k\simeq1+o_{k2}(\gamma_2)+o_{k3}(\gamma_3)+...$ and $\alpha_k\simeq1+o'_{k2}(\gamma_2)+o'_{k3}(\gamma_3)+...$ (here $k$ is the number of iterations and $o_{k2,k3..}(x)$, $o'_{k2,k3..}(x)$ mean small values in order of argument). We define the parameters $\alpha_{k\to\infty}$ and $\beta_{k\to\infty}$ from the iteration procedure for the special case considered the Sec. IV. The obtained non-homogeneous equation is simply  resolved
 \begin{eqnarray}
\varphi^{L}_1(x)=\frac{\exp(-x)}x+\sum_{l=2}^{L}\frac{\gamma_{l}\beta_0^{l}}{2x}\int_x^{\infty}\frac{\{\exp[-x-x'(\alpha_0l-1)]-\exp[x-x'(\alpha_0l+1)]\}}{x'^{l-1}} dx',\label{eq4}
 \end{eqnarray}
where upper index $L$ identify the number of taken into account nonlinear terms and lower index identify the number of iterations was done. It is easy to see that for $L>2$ iteration procedure is divergent in the region $x<x_0(L)$, because $|\varphi^{L}_{k+1}(x)/\varphi^{L}_k(x)|_{x\to+0}\to\infty$. On the other hand, relaxation processes are introduced through the Fourier components $\Phi(q)$ of the potential in Born approximation, and the region of divergence $x<x_0(L)$ may became are not essential for them for the certain range of parameters. The Fourier transformation of the eq.(\ref{eq4}) is possible for $L\leq4$ and has the next form
 \begin{eqnarray}
\Phi^{L}_1(q)=\frac{4\pi}{q^2+1}+\sum_{l=2}^{L}\gamma_{l}\beta_0^{l}F_{l}(q,\alpha_0),\label{eq5}
 \end{eqnarray}
where $q={\rm q}\lambda$ and ${\rm q}$ is the wave-vector transfer value. To obtain usual dimension one has to multiply the Fourier component by the factor  $Z|e|\lambda^2/\epsilon$. The coefficients $F_{l}(q)$ are defined as the next
 \begin{eqnarray}
F_{l}(q,\alpha_0)=\frac{4\pi}{q}\times\left\{
\begin{array}{lll}
  \frac1{1+q^2}\left[q\ln\left(\frac{2\alpha_0+1}{2\alpha_0-1}\right)-2\arctan\left(\frac{q}{2\alpha_0}\right)\right],&l=2; \\
\frac1{1+q^2}\left\{6\alpha_0\arctan\left(\frac{q}{3\alpha_0}\right)+q\ln|9\alpha_0^2+q^2|-q\ln\left[\frac{(3\alpha_0+1)^{3\alpha_0+1}}{(3\alpha_0-1)^{3\alpha_0-1}}\right]\right\},&l=3 ;\\
\arctan\left(\frac{q}{4\alpha_0}\right)\left(\frac{q^2-16\alpha_0^2}{1+q^2}\right)+\frac{q}{1+q^2}\left\{\frac12\ln\left[\frac{(4\alpha_0+1)^{(4\alpha_0+1)^2}}{(4\alpha_0-1)^{(4\alpha_0-1)^2}}\right]-4\alpha_0\ln|16\alpha_0^2+q^2|\right\},&l=4 ;
\end{array}
\right.\label{eq6}
 \end{eqnarray}
One can see from eq. (\ref{eq5}) that for any $q<q_{max}$ coefficients $|F_{l}(q)|<const(q_{max})$. Therefore, if ${\gamma_{l}}$ are small enough, then eqs. (\ref{eq4}) and (\ref{eq5}) with $\beta_0=1$ and $\alpha_0=1$ give valid additives to the Fourier components of the Brooks-Herring screened potential.
Next, we consider bulk GaAs as an example. To calculate $\{\gamma_l\}$ we need the equation of electroneutrality
 \begin{eqnarray}
n(\zeta,T)=\left\{
\begin{array}{lll}
N_d/\{\exp[(\zeta-E_d)/T]+1\},&N_d<10^{16}\mbox{cm}^{-3}; \\
N_d,&N_d\geq10^{16}\mbox{cm}^{-3}; \\
\end{array}
\right.\label{eq7}
 \end{eqnarray}
where $N_d$ is the donor concentration and $E_d$ is the donor energy level. In the latter formula we suppose that donor concentration higher than $10^{16}$ cm$^{-3}$ leads to the Mott transition in bulk GaAs (see for example pg. 41, Fig. 21 in ref. \cite{Brillson}). Figures \ref{Fig1} (a-c) demonstrate coefficients $\gamma_{2-4}$  for n-GaAs parameters \cite{GaAs_parameters} versus temperature $T$ and donor concentration $N_d$ as the contours of equivalent values. The calculations performed for $Z=1$. Obviously, in the case $Z=-1$ (acceptors) $\gamma_{2,4}$ change their signs to opposite [see eq.(\ref{eq3})]. Therefore, screened potential loose its symmetry relative to the sign of the centers charge, when the nonlinear terms are taken into account.

\begin{figure}
\centering
\includegraphics[width=3in]{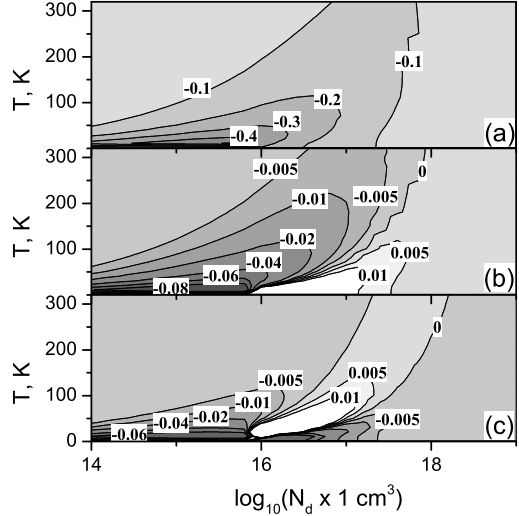}
\caption{The relative additives to mobility in n-GaAs versus temperature $T$ and donor concentration $N_d$ (logarithmic scale) as the contour plots. Frames a-c correspond to $\Upsilon_{2-4}$. \label{Fig2}}
\end{figure}

\section{Additives to mobility}
In the present section we calculate the nonlinear additives to the ionized impurity scattering mobility in n-GaAs\cite{GaAs_parameters}. The transport time for the electron-impurity scattering can be calculated according to the formula\cite{Gantmakher&Levinson,Seeger}
 \begin{eqnarray}
\tau_L^{-1}(E)=\frac{N_d(Ze^2)^2}{8\pi\sqrt{2m}\epsilon^2E^{3/2}}\int_0^{2\sqrt{2mE}\lambda/\hbar}dqq^3|\Phi_1^L(q)|^2,\label{eq8}
 \end{eqnarray}
where $E$ is the electron energy and $\Phi_1^L(q)$ is defined by the formulas (\ref{eq5}), and (\ref{eq6}) with $\alpha_0=1$ and $\beta_0=1$. The low-field mobility of the electron gas can be introduced in the form \cite{Seeger}:
 \begin{eqnarray}
\mu_{imp}^{L}=\frac{16\pi\sqrt{2m}T^{3/2}|e|}{3n(\zeta,T)(2\pi\hbar)^3}\int_0^{\infty}dzz^{3/2}\tau_L(zT)~\frac{\exp(z-\zeta/T)}{[\exp(z-\zeta/T)+1]^2}.\label{eq9}
 \end{eqnarray}
 To demonstrate the influence of the non-linear additives on mobility we introduce the next dimensionless parameter
 \begin{eqnarray}
\Upsilon_L=\frac{\mu_{imp}^{L}-\mu_{imp}^{L-1}}{\mu_{imp}^{1}},\label{eq10}
 \end{eqnarray}
where $\mu_{imp}^{1}$ is the mobility calculated using the Brooks-Herring approach. The parameters $\{\Upsilon_L\}$ may be treated as the relative mobility changes due to the $L$-th non-linear term. The results of calculation for GaAs parameters one can see in Fig. \ref{Fig2} (a-c).

It is easy to see that outside the region $T <50K$, $N_d \in (10^{14}\mbox{cm}^{-3},2\times10^{16}\mbox{cm}^{-3})$ non-linear additives change the mobility no more then 20\%. Whereas, inside the region $T <50K$, $N_d \in (10^{14}\mbox{cm}^{-3},2\times10^{16}\mbox{cm}^{-3})$  the changes in mobility with the addition of the quadratic term $\Upsilon_2$ is very significant see [Fig. \ref{Fig2} (a)]. On the other hand, with increasing amounts of non-linear terms from $L=2$ to $L=4$, the mobility does not change substantially [see Figs. \ref{Fig2} (b) and (c)]. Correspondingly, one may expect that for $L=2$ the mean inaccuracy of the formulas (\ref{eq5}) and (\ref{eq6}) caused by the power-series cutoff at most 10\% in low the temperature region and below 5\% for $T>50K$ [see Fig. 1(b)]. In the case $L=3$ inaccuracy decreases further more [see Fig. \ref{Fig2}(c)].

\section{Long-wave limit}
Considering eqs. (\ref{eq5}) and (\ref{eq6}), one can see that  for $q\ll1$ Fourier components in eq. (\ref{eq5}) can be rewritten as
\begin{eqnarray}
\beta_1-\beta_1\alpha_1^2q^2=1+\sum_{l=2}^{L}\frac{\gamma_{l}\beta_0^{l}}{4\pi}F_{l}(0,\alpha_0)-\left[1-\left.\sum_{l=2}^{L}\frac{\gamma_{l}\beta_0^{l}}{8\pi}\frac{\partial^2 F_{l}(q,\alpha_0)}{\partial q^2}\right|_{q=0}\right]q^2+O_1(q^4),\label{eq11}
 \end{eqnarray}
where we suppose that $\Phi^{L}_1(q)=4\pi\beta_1/(1+\alpha_1^2q^2)+O_2(q^4)=4\pi\beta_1(1-\alpha_1^2q^2)+O_3(q^4)$ and the functions $O_{1-3}(x)$ describe small values in order of the argument. It is easy examining that functions $\Phi^{L}_k(q)\simeq4\pi\beta_k/(1+\alpha_k^2q^2)$ conserve this form for arbitrary number of iterations $k$.  Neglecting the small values $O_{1-3}(q^4)$ and supposing that the number of convergent iterations is infinite, one could obtain the algebraic system of equations for the parameters $\alpha=\alpha_{N\to\infty}$ and $\beta=\beta_{N\to\infty}$:
 \begin{eqnarray}
\left\{
\begin{array}{lll}
\sum_{l=2}^{L}\gamma_{l}\beta^{l}a_l(\alpha)-\beta+1=0;\\
\sum_{l=2}^{L}\gamma_{l}\beta^{l}b_l(\alpha)+\alpha^2\beta-1=0;\\
\end{array}\right.\label{eq12}
 \end{eqnarray}
where the coefficients $a_{2,3,4}(\alpha)=F_{2,3,4}(0,\alpha)/4\pi$ are defined by the equality
  \begin{eqnarray}
a_{l}(\alpha)=\left\{
\begin{array}{lll}
  \ln\left(\frac{2\alpha+1}{2\alpha-1}\right)-\frac{1}{\alpha},&l=2; \\
2+2\ln|3\alpha|-\ln\left[\frac{(3\alpha+1)^{3\alpha+1}}{(3\alpha-1)^{3\alpha-1}}\right],&l=3 ;\\
-4\alpha-8\alpha\ln|4\alpha|+\frac12\ln\left[\frac{(4\alpha+1)^{(4\alpha+1)^2}}{(4\alpha-1)^{(4\alpha-1)^2}}\right],&l=4 ;
\end{array}
\right.\label{eq13}
 \end{eqnarray}
and the coefficients $b_{2,3,4}(\alpha)=(8\pi)^{-1}\left.\partial^2 F_{2,3,4}(q,\alpha)/\partial q^2\right|_{q=0} $ are defined by the next equality
  \begin{eqnarray}
b_{l}(\alpha)=\left\{
\begin{array}{lll}
 \frac1{12\alpha^3}+\frac1{\alpha}-\ln\left(\frac{2\alpha+1}{2\alpha-1}\right),&l=2; \\
\frac{1}{27\alpha^2}+\ln\left[\frac{(3\alpha+1)^{3\alpha+1}}{(3\alpha-1)^{3\alpha-1}}\right]-2-2\ln|3\alpha|,&l=3 ;\\
\frac1{12\alpha}-\frac12\ln\left[\frac{(4\alpha+1)^{(4\alpha+1)^2}}{(4\alpha-1)^{(4\alpha-1)^2}}\right]+4\alpha+4\alpha\ln|16\alpha^2|,&l=4 .
\end{array}
\right.\label{eq14}
 \end{eqnarray}
If the algebraic system (\ref{eq12}) with given $L$ and coefficients (\ref{eq13}), (\ref{eq14}) has real positive solutions then the iteration procedure is supposed to be convergent. Choosing the solution that is closest to the pair $\alpha=1$ and $\beta=1$ one can use the Fourier components
 \begin{eqnarray}
\Phi^{L}_{N\to\infty}(q)\approx\frac{4\pi\beta}{1+\alpha^2q^2},\qquad q\ll1.\label{eq15}
 \end{eqnarray}
The latter approach valid for processes with a small wave-vector transfer $q$, but for arbitrary $\gamma_{2,3,4}$ that provide eqs. (\ref{eq12})-(\ref{eq14}) with real positive roots. To analyze the system we consider limiting case $q\to0$ and put $L=2$. Under such conditions we obtain $\alpha=1$ and only one equation $0.1\gamma_2\beta^2-\beta+1=0$. The real root $\beta=5[1-\sqrt{1-0.4\gamma_2}]/\gamma_2$ that satisfies the mentioned above conditions exists only if $\gamma_2<2.5$. Note, that in the case $Z<0$ the latter inequality is always  satisfied. On the other hand, in the case $Z>2$ it fails in the low temperature region [see Fig. \ref{Fig1}(a)]. The linearization of the system (\ref{eq12}) leads to the approximate solution $\alpha\simeq1-0.04\gamma_2$ and $\beta\simeq1+0.1\gamma_2$. Therefore, one can expect that if $Z>0$ then $\alpha<1$ and $\beta>1$, whereas if $Z<0$ then $\alpha>1$ and $\beta<1$. One also can estimate that inaccuracy of the first iteration, considered in Sec. III is about $0.1\gamma_2\lesssim10$\% for slow particles [see Fig. \ref{Fig1}(a)].

 The physical interpretation of the approach (\ref{eq12})--(\ref{eq15}) is simple:  the Brooks-Herring screened potential remains valid for slow particles, but the screening length became in $\alpha$ times larger and the charge of the ionized impurity became $\beta/\alpha^2$ times larger.

\section{conclusion}
The validity of the obtained results restricted by the conditions for the Born and the effective mass approximations. The latter supposes $\lambda\gg l_c$ ($l_c$ is the lattice constant) that is well satisfied for the considered doping levels, whereas Born approximation is questionable for some regions of parameters. We also omit consideration of the nonparabolicity effect that is not essential for the investigation performed.

The main theoretical results are next: (i) the screened potential loose its symmetry relative to the sign of the scattering centers charge, when the nonlinear terms are taken into account; (ii) in the region $T <50K$, $N_d \in (10^{14}\mbox{cm}^{-3},2\times10^{16}\mbox{cm}^{-3})$  the nonlinear terms changes the ionized impurity scattering mobility up to 50\% in n-GaAs; (iii) the Brooks-Herring screened potential is valid in low temperature region under the condition $q\equiv{\rm q}\lambda\ll1$, but with the different screening length $\lambda'=\alpha\lambda$  and the different charge of the scattering center $\beta Z|e|/\alpha^2$, where $\alpha\simeq1-0.04\gamma_2$, $\beta\simeq1+0.1\gamma_2$ and $\gamma_2=(Ze^2/2\epsilon\lambda)[{\partial^2 n(\zeta,T)}/{\partial \zeta^2}][{\partial n(\zeta,T)}/{\partial \zeta}]^{-1}$.

\section*{References}

\begin{thebibliography}{100}
\bibitem{GaAs_parameters} GaAs parameters: effective mass is $m=0.067m_0$,  static dielectric constants is $\epsilon=12.9$.
\end{thebibliography}

\end{document}